\documentclass[%
 jmp,
 amsmath,amssymb,
preprint,%
]{revtex4-1}

\draft 

\usepackage{amsmath,amssymb,amsfonts,amscd,amsxtra}
\usepackage{latexsym}
\usepackage{bm}
\usepackage{tikz-cd}
\usepackage{pgfplots}

\input{epsf.sty}
\usepackage{epsfig}
\usepackage{graphicx,color}
\usepackage{graphics}
\usepackage{subcaption}
\usepackage{appendix}
\usepackage{wrapfig, blindtext}
\usepackage[top=1in, bottom=1in, left=1in, right=1in]{geometry}
\usepackage{titlesec}
\titlelabel{\thetitle\quad}

\newtheorem{thm}{Theorem}[section]

\newtheorem{lem}[thm]{Lemma}
\newtheorem{prop}[thm]{Proposition}
\newtheorem{defn}{Definition}
\newtheorem{rem}{Remark}

\newtheorem{example}[thm]{Example}

\renewcommand{\thesection}{\arabic{section}}
\renewcommand{\theequation}{\thesection.\arabic{equation}}

\newcommand{\be}{\begin{equation}}
\newcommand{\ee}{\end{equation}}

\newcommand{\R}{\mathbf{R}}

\newcommand{\B}{\mathcal{B}}

\newcommand{\Lo}{\mathcal{L}}

\newcommand{\qed}{\hfill \ensuremath{\square}}
\renewcommand\appendix{\par
  \setcounter{section}{0}
  \setcounter{subsection}{0}
  \setcounter{figure}{0}
  \setcounter{table}{0}
  \renewcommand\thesection{Appendix \Alph{section}}
  \renewcommand\theequation{\Alph{section}.\arabic{equation}}
  \renewcommand\thefigure{\Alph{section}.\arabic{figure}}
  \renewcommand\thetable{\Alph{section}.\arabic{table}}
  \renewcommand\thethm{\Alph{section}.\arabic{thm}}
}

\newcommand{\CC}{\mathbf{C}}
\newcommand{\NN}{\mathbf{N}}
\newcommand{\RR}{\mathbf{R}}

\newcommand{\ZZ}{\mathbf{Z}}

\numberwithin{equation}{section}

\usepackage[pagebackref=false,colorlinks]{hyperref}

\begin{document}


\title{Bulk-interface correspondences for one dimensional topological materials with inversion symmetry}  



\author{Guo Chuan Thiang}
\email[]{guochuanthiang@bicmr.pku.edu.cn}
\thanks{}
\affiliation{Beijing International Center for Mathematical Research, Peking University, Beijing, China}

\author{Hai Zhang}
\email[]{haizhang@ust.hk}
\thanks{}
\affiliation{Department of Mathematics, Hong Kong University of Science and Technology, Hong Kong S.A.R., China}



\begin{abstract}
The interface between two materials described by spectrally gapped Hamiltonians is expected to host an in-gap interface mode, whenever a certain topological invariant changes across the interface. We provide a precise statement of this bulk-interface correspondence, and its rigorous justification. The correspondence applies to continuum and lattice models of interfaces between one-dimensional materials with inversion symmetry, with dislocation models being of particular interest. For continuum models, the analysis of the parity of the ``edge'' Bloch modes is the key component in our argument, while for the lattice models, the relative Zak phase and index theory are.

\end{abstract}


\maketitle 



\setcounter{equation}{0}
\setlength{\arraycolsep}{0.2em}

\section{Introduction}
In this paper, we study the phenomenon of \emph{bulk-interface correspondence}, whereby the interface between two distinct spectrally-gapped one-dimensional materials necessarily hosts a robust localized mode in the spectral gap. This is analyzed in both lattice and continuum models. Such modes arise from the interplay between two ``distinct phases'' pieced together along an interface. They have many applications in the localization and transportation of wave energy and are extensively studied in the fields of photonics and phononics \cite{Joanno-11, Ozawa-19}.

The prototype was introduced by Su--Schreiffer--Heeger in their seminal work on polyacetylene polymer chains \cite{ssh-79}, in which an alternating sequence of single bonds and double bonds occurs on one side of a domain wall, while the reverse sequence occurs on the other side (see Eq.\ \eqref{eqn:SSH.lattice} for a sketch). The similarly influential domain-wall Dirac Hamiltonian, considered in \cite{jr-76,js-81}, has a ``mass term'' with differing signs on either side of the wall. Both were argued to have zero-energy localized solutions, without explicit use of topology or index theory ideas.

From a modern perspective, we may label the possible ``bulk phases'' of a 1D material by suitable topological invariants (e.g.\ a quantized Zak phase \cite{Zak-89}, or the sign of a nonzero mass term). A subtle but crucial point is that only the \emph{difference} of such invariants has unambiguous meaning \cite{thiang-15}, so it is the manner in which two systems are pieced together along an interface which determines the unambiguous ``topological non-triviality'' of the combined system. For example, in generic \emph{dislocation} models, such non-triviality is automatic from our Lemma \ref{lem:phase.shift}, and will generally lead to an interface mode (Prop.\ \ref{prop:dislocation.mode}, Theorem \ref{thm:dislocated.bulk.interface.correspondence}); no explicit calculation of the rather abstract (relative) topological invariant on either side of the interface is necessary. 

Indeed, in \cite{ssh-79,jr-76,js-81}, one had two possible degenerate ground states, neither of which is preferred, but one or the other must be locally picked out under spontaneous breaking of a \emph{reflection} symmetry. Likewise, for half-space continuum models, reflection symmetry and the \emph{choice} of reflection plane for the boundary termination (for the same bulk Hamiltonian), play key roles for the existence of edge states, as has been known for a long time \cite{Zak-85}.

In recent literature, the bulk-\emph{edge} correspondence for \emph{two}-dimensional materials occupying a \emph{half-space} has gained in mathematical precision and generality. For lattice models, see, e.g., \cite{hatsugai-93,Graf-02,gp-13}, as well as \cite{KRS-02,PSB-16,BSS} for the disordered case, and \cite{thiang-20} for general boundary geometries; For continuum models, see, e.g., \cite{bal-19} for Dirac Hamiltonians, \cite{combes-05} for quantum Hall Hamiltonians, and \cite{KLT-22} for general Riemannian surfaces. To obtain a similar result for \emph{one}-dimensional half-space lattice models, one has to impose a very restrictive ``chiral/sublattice symmetry'' assumption; see, e.g., \cite{mong-11} and \S 2.3 of \cite{PSB-16}. This latter correspondence is really the classical index theorem for Toeplitz operators in disguise, recalled in Section \ref{thm:BEC.chiral}. 

The interface mode problem, which is our actual focus, requires a more careful treatment of ``topological invariants''. Once clarified, we obtain a quick index-theoretic proof that the SSH \emph{interface} model hosts an interface mode (Prop.\ \ref{prop:SSH}), \emph{provided chiral symmetry holds}. The interface mode of the domain-wall Dirac Hamiltonian can also be deduced from the Callias index formula \cite{callias-78}. However, these methods may be unsatisfactory because actual materials are more realistically modelled with differential operators \emph{without} strict chiral symmetry. In this setting, the nature of their interface modes has recently attracted mathematical attention, e.g., \cite{Fefferman-Lee-Thorp-Weinstein-17,druout-20-1}. See also \cite{ammari-20-1, ammari-20-3} for phononic materials made of high contrast resonating bubbles.

\textbf{Outline and main results.} This paper is split into two main parts, which may be read independently and in any order. The first part, Section \ref{sec:inv_sym}, concerns inversion symmetric \emph{continuum} models, Eq.\ \eqref{eq-photonic}, describing photonic structures (these are defined on real function spaces). 
If there is a spectral gap, we can define a $\pm$-valued bulk index, closely related to a quantized Zak phase. Our main result is that the interface of two gapped materials with different bulk indices will have exactly one in-gap interface mode (Theorem \ref{thm-existence_int_mode}). This constitutes a \emph{bulk-interface correspondence}.

The second part, Section \ref{sec:discrete.models}, concerns inversion symmetric \emph{lattice} models, defined over the reals. We explain how an auxiliary sublattice operator appears, and therefore a notion of \emph{approximate} chiral symmetry. In Section \ref{sec:relation.between.invariants}, we clarify the relationship between various bulk indices used in the literature, and focus on the quantized Zak phase. As explained in Section \ref{sec:unit.cell.convention}, only \emph{relative} Zak phases have well-defined meanings, and this motivates the interface model in Section \ref{sec:dislocation.model}, which has the SSH model as a special case. Our main results are: a bulk-edge correspondence, Theorem \ref{thm:PT.BEC}, and a bulk-interface correspondence, Theorem \ref{thm:dislocated.bulk.interface.correspondence}. These hold as long as the strictly nearest-neighbour terms dominate, but may be false otherwise. 

Finally, in Section \ref{sec:comparison}, we discuss some differences between continuum and lattice models, as well as the outlook for future work.

\section{One dimensional continuum models with inversion symmetry}\label{sec:inv_sym}
In this section, we investigate one-dimensional topological structures with inversion symmetry. We shall restrict to photonic/phononic systems; the extension to electronic systems is straightforward and will be discussed at the end of this Section.
The corresponding periodic differential operator is given by
\begin{equation} \label{eq-photonic}
    \Lo\psi = -\frac{1}{\varepsilon(x)} \dfrac{d}{dx} \left( \frac{1}{\mu(x)}\frac{d\psi}{dx}\right) \quad\mbox{for} \; x\in\mathbf{R},
    \end{equation}
and the coefficients $\varepsilon,\mu$ satisfy the following two conditions:
\begin{itemize}
\item 
 The permittivity $\varepsilon(x)$ and the permeability $\mu(x)$ are piecewise continuous positive real-valued functions with period one:
\begin{equation*}
    \varepsilon(x)= \varepsilon(x+1), \quad    \mu(x)= \mu(x+1).
\end{equation*}

\item
$
\mathcal{P}\Lo= \Lo \mathcal{P}, 
$
where $\mathcal{P}$ is the \textbf{parity operator} defined by 
$$
\mathcal{P} \psi(x) = \psi(-x)
$$
for any function $\psi:\RR\to \CC$.
\end{itemize}

Under the above assumptions, we see that $\varepsilon(x)= \varepsilon(-x)$, $\mu(x) = \mu(-x)$, or equivalently, 
$\varepsilon(x)= \varepsilon(1-x)$, $\mu(x) = \mu(1-x)$. Also, $\Lo$ is \textbf{time-reversal symmetric} in the sense that it commutes with the operation of complex conjugation.

Such operators were investigated in \cite{chan-14} and \cite{lin-zhang-21}. It was shown that a localized mode exists at the interface of two semi-infinite periodic structures with different bulk topological indices. We shall improve the argument in \cite{lin-zhang-21} and derive a stronger bulk-interface correspondence result that is able to characterize precisely the number of interface modes.  The new argument is self-contained, and does not rely on the transfer matrix technique or the oscillatory theory of Sturm--Liouville systems.

\subsection{Preliminaries}
We first recall some facts about the spectrum of the periodic ordinary differential operator $\Lo$, and the regularity of its solutions.
For each real-valued $E$, by the standard regularity theory of ODEs,
we know that the solutions to the equation $(\Lo-E)u=0$ are absolutely continuous in $\mathbf{R}$. Moreover, the function $\frac{u'(x)}{\mu(x)}$ is also absolutely continuous. Here and throughout, we use the notation $u'(\cdot)$ to denote the derivative of $u$ with respect to the variable $x$.  At a point of discontinuity of $\mu$, say $x=x_0$,  we interpret the value $\frac{1}{\mu(x_0)}u'(x_0)$ as either 
the left-sided limit $\lim_{x \to x_0^-}\frac{1}{\mu(x)}u'(x)$ or the right-sided limit $\frac{1}\lim_{x \to x_0^+}\frac{1}{\mu(x)}u'(x)$.
The two one-sided limits are equal by the regularity of the solution $u(x)$. For ease of notation, we use the notation 
$\frac{1}{\mu(x_0)}u'(x_0)$ for either of the two one-sided limits in subsequent analysis.

The spectrum of the operator $\Lo$ can be analyzed using the standard Floquet--Bloch theory. 
Let $\B = [-\pi, \pi]$ be the Brillouin zone and $[0, \pi]$ be the reduced Brillouin zone. Denote by $AC[0, 1]$ the space of absolutely continuous functions defined on $[0, 1]$.  For each Bloch wavenumber $k \in \B$, we consider the following 
one-parameter family of Floquet--Bloch
eigenvalue problems,  
\begin{equation} \label{eigen-k}
     \Lo \psi(x)= E \psi(x) \quad x\in [0, 1], 
\end{equation}
in the function space
$$
V_k = \left\{u \in AC[0, 1]: u(1)= e^{ik} u(0), \frac{u'}{\mu} \in AC[0, 1], \frac{u'(1)}{\mu(1)}= e^{ik}\frac{u'(0)}{\mu(0)}  \right\}. 
$$
equipped with the following inner product
\begin{equation}\label{eq:inner_prod}
(u, v)=\int_{0}^1 \varepsilon(x) \bar{u}(x) {v}(x) dx.
\end{equation}
Here and throughout, $\bar{u}(x)$ denotes the complex conjugate of $u(x)$. 
It is easy to check that 
for each $k\in\B$, the eigenvalue problem (\ref{eigen-k}) is self-adjoint and attains a discrete set of real eigenvalues with finite multiplicity,
$$
E_1(k) \leq E_2(k) \leq \cdots \leq E_j(k) \leq \cdots. 
$$
We have the following properties of the
function $E_j(k)$, also called the dispersion relation of the $j$-th spectral band.  See \cite{Reedsimon, lin-zhang-21} for proof. 

\begin{lem} \label{lem-11}
\begin{enumerate}
\item [(1)]
The function $E_j(k)$ is Lipschitz continuous with respect to $k\in \B$. 

\item [(2)]
$E_j(k) = E_j(-k)$ holds for each $k\in \B$. Moreover, $E_j(k)$ can be extended to a periodic function in $k$ with period $2\pi$, i.e.  $E_j(k)= E_j(k+ 2\pi)$.

\item  [(3)]
$E_j(k)$ are strictly monotonic on each of the half Brillouin zones $(-\pi, 0)$ and $(0, \pi)$.

\end{enumerate}
\end{lem}

For each $j\geq 1$, we define the \textbf{band edges} to be
$$
E_j^-=\min \{ E_j(k): k\in \B \}, \quad  E_j^+=\max \{ E_j(k): k\in \B\}. 
$$
Then the entire spectrum of the operator $\Lo$ on $L^2(\R)$ is given by
$$
\sigma(\Lo) = \bigcup\limits_{j\geq 1} \, [E_j^-, E_j^+],
$$
and corresponds to the essential spectrum of the operator. Moreover, $\Lo$ has no point spectrum.

If $E_j^+ < E_{j+1}^-$, the spectrum contains a band gap between the $j$-th and $(j+1)$-th bands. Note that by the monotonicity in Lemma \ref{lem-11}, the band edges $E_j^+, E_j^-$ occur at either $k=0$ or $k=\pi$. Moreover, we have the following result. See Theorem 2.5 in \cite{lin-zhang-21} for a proof. 

\begin{lem} \label{lem-11-2}  
For each $j \geq 1$, we have either
$$
E_j^+ = E_j(0), \,\,  E_{j+1}^-= E_{j+1}(0),
$$
or
$$
E_j^+ = E_j(\pi), \,\,  E_{j+1}^- =E_{j+1}(\pi).
$$
\end{lem}

Note that the eigenfunction associated with the eigenvalue $E_j(k)$ can be extended to a function on $\RR$, through the following formula
$$
u(x+1) = u(x) e^{ikx}.
$$
Moreover, the extended function still satisfies the eigen-equation 
$\Lo \psi(x)= E \psi(x)$ for all $x\in \mathbf{R}$. 
This extended function is called a $j$-th Bloch eigenfunction (or Bloch mode) and is denoted $\varphi_{j, k}$. For ease of notation, we shall use the same symbol for both the $L^2[0,1]$-normalizable function, and its (non-normalizable) extended version.

\begin{defn}
We say that a Bloch mode $\varphi_{j,k}$ has even-parity (odd-parity) if $\varphi_{j,k}$ is an even (odd) function.
\end{defn}

\begin{lem} \label{lem-blochmode}
Let $\Lo$ be a periodic operator of the form (\ref{eq-photonic}). Let $E^*$ be an eigenvalue of the Floquet--Bloch eigenvalue problem of $\Lo$ in $V_{k=k^*}$ where $k^*=0,$ or $\pi$. 
Then the following hold:

\begin{enumerate}
    \item The space of Bloch eigenfunctions in $V_{k=k^*}$ associated with $E^*$ has dimension at most two. Moreover, the eigenbasis  can be chosen to be real-valued functions. 
    \item 
    If the above-mentioned dimension is one, then the eigenspace is spanned by a real-valued function which can be chosen to be either even or odd. 
    \item  
    If the above-mentioned dimension is two, then the eigenspace is spanned by two real-valued functions, one of which is even while the other is odd.  
\end{enumerate}
\end{lem}
\noindent\textbf{Proof.} We only prove the Lemma for the case $k^*=0$. The case $k^*=\pi$ can be proved similarly. 

Proof of (1). We consider the solutions to the second order ordinary differential equation $\Lo \psi = E^*\psi$. It is clear that a solution $\psi$ is uniquely determined by $\psi(0)$ and $\psi'(0)$. Therefore, the space of solutions has dimension at most two. On the other hand, note that if $\psi \in V_{k=0}$ solves $\Lo \psi = E^*\psi$, then so do the real and imaginary parts of $\psi$. Therefore, the eigenbasis can be chosen to be real-valued functions. 

Proof of (2).  By (1), we can choose a real-valued eigenfunction $\psi$ that spans the space of Bloch eigenfunctions in $V_{k=0}$ associated with $E^*$. Since $\Lo$ is inversion symmetric, it is easy to check that $\mathcal{P}\psi$ is also a real-valued eigenfunction in the space of Bloch eigenfunctions in $V_{k=0}$ associated with $E^*$. Therefore, we have $\mathcal{P}\psi = \pm \psi $, from which the claim in (2) follows. 

Proof of (3). We first show that the Bloch eigenfunctions of $\Lo$ in $V_{k=0}$ associated with the eigenvalue $E^*$ cannot all be even. Otherwise, all the eigenfuntions have vanishing Neumann data and hence are linearly dependent (using the same argument as in (1)). Similarly, the Bloch eigenfunctions cannot all be odd. On the other hand, due to the inversion symmetry of the operator $\Lo$, the Bloch eigenfunctions can be chosen to be either even or odd. Therefore, we can choose two real-valued functions, with one even and the other odd, such that they span the Bloch eigenspace.  \qed

Motivated by the above Lemma, we introduce the following subspaces of $V_{k=k^*}$ with $k^*=0$ or $\pi$:
$$
V_{k=k^*, e} = \{f \in V_{k=k^*}: f= \mathcal{P}f\}, \quad V_{k=k^*, o} = \{f \in V_{k=k^*}: f =-\mathcal{P}f\}.
$$
It is clear that $V_{k=k^*}$ is an orthogonal sum of the two subspaces $V_{k=k^*, e}, V_{k=k^*, o}$. Moreover, the above Lemma implies that the spectrum of $\Lo$ restricted to $V_{k=k^*}$ is the union of the spectrum of $\Lo$ restricted to the two subspaces.

We are now ready to investigate the change of parity for the Bloch modes at the two extremal points in a band gap. See also \cite{lin-zhang-21} for a different proof using the  oscillation theory for Sturm--Liouville operators. 

\begin{prop} \label{thm-parity_change}
Let $\Lo$ be a periodic operator of the form (\ref{eq-photonic}). Assume that there is a band gap between the $j$-th and $(j+1)$-th bands. Then the Bloch modes at $(k^*, E_j^+)$ and at $(k^*,E_{j+1}^-)$ have different parities, where $k^*=0$ or $\pi$.
\end{prop}
\noindent\textbf{Proof.} 
Without loss of generality, we only prove the case $k^*=0$, where the maximum of the $j$-th band and the minimum of the $(j+1)$-th band are attained at $k=0$.  

Step 1. We apply a continuous family of perturbations to the operator $\Lo$ such that both time-reversal symmetry and inversion symmetry are preserved, and that the band gap between the $j$-th and $(j+1)$-th bands can be closed. 
This can be done by considering  the following family of operators $\Lo_{s}$ with coefficients 
$$
\varepsilon_s(x)= \varepsilon(x) + s(1- \varepsilon(x)),  \quad  \mu_s(x) = \mu(x)+ s(1-\mu(x)),\qquad 0\leq s\leq 1.
$$
We denote by $s_1$, the first value of $s$ such that the $j$-th band gap closes.
It is clear that $0<s_1 \leq 1$. 

Step 2. We consider the operator $\Lo_{s_1}$. Using Lemma \ref{lem-11-2}, we can deduce that the maximum of the $j$-th band and the minimum of the $(j+1)$-th band the family of operators $\Lo_{s}$ are always attained at $k=0$. Let $(k=0, E^*)$
be the touching point of the $j$-th band and $(j+1)$-th band of $\Lo_{s_1}$. By Lemma \ref{lem-blochmode},  $E^*$ is an eigenvalue of $\Lo_{s_1}$ in both subspaces $V_{k=k^*, e}$ and  $V_{k=k^*, o}$. 
By Lemma \ref{lem-blochmode} again, we see that $E^*$ is a non-degenerate eigenvalue for $\Lo_{s_1}$ in both subspaces $V_{k=k^*, e}$ and $V_{k=k^*, o}$. 

Step 3. We consider the eigenvalue problem of $\Lo_{s}$ in the subspaces $V_{k=k^*, e}$ and $V_{k=k^*, o}$ for $s<s_1$ but close to $s_1$. Using standard perturbation theory for eigenvalues of self-adjoint operators, we see that $\Lo_{s}$ has two eigenvalues in $V_{k=k^*}$; one is 
perturbed from $E^*$ in the space $V_{k=k^*, e}$, and the other from the space $V_{k=k^*, o}$. Note that the $j$-th band gap of $\Lo_{s}$ is open for $s<s_1$. We see that for all $s<s_1$ and sufficiently close to $s_1$, the $j$-th and $(j+1)$-th Bloch modes of $\Lo_{s}$ at $k=k^*$ have different parities. 

Step 4. Finally, notice that for all $0<s<s_1$, the parity of the $j$-th Bloch mode at $k=k^*$ remains the same, and similarly for the $(j+1)$-th Bloch mode. We conclude that the 
$j$-th and $(j+1)$-th Bloch modes of $\Lo$ at $k=k^*$ have different parities. 
\qed

\subsection{Bulk  topological phases under inversion symmetry}
Let $\Lo$ be an inversion symmetric periodic operator of the form (\ref{eq-photonic}). 
Assume that there is a gap between the $j$-th and $(j+1)$-th spectral bands of $\Lo$, and let $E$ be a real number in the band gap. Recall that the lower edge of the band gap, i.e. the maximum of the $j$-th band, $E_j^+$, is achieved at either $k=0$ or $\pi$. With respect to this band gap, we define the following bulk index: 
$$
\gamma_j =: \mbox{ the parity of the Bloch mode at $E_j^+$}. 
$$
 
For $\Lo$ of the form (\ref{eq-photonic}), we observe that the lowest Bloch eigenvalue is zero and is attained at $k=0$ with a constant, thus even, Bloch function. In the event that every spectral band below the $j$-th one is isolated, we can show that 
\begin{equation*}\label{eq-zeta}
\gamma_j=(-1)^{j-1} e^{i \sum_{m=1}^{j} \theta_m},
\end{equation*}
where $\theta_m$ is the Zak phase for the $m$-th isolated band. As explained at the end of Section \ref{sec:relation.between.invariants}, the Zak phase has the following equivalent definition:
$$
\theta_m = \begin{cases}
0, \quad \mbox{if $\varphi_{m,0}(x) \; \mbox{and} \; \varphi_{m,\pi}(x)$ attain the same parity}, \\ 
\pi, \quad \mbox{if $\varphi_{m,0}(x) \;  \mbox{and} \; \varphi_{m,\pi}(x)$ attain different parities}.
\end{cases}
$$
For the more involved case where the bands below the spectral band gap may cross each other, we refer to \cite{lin-zhang-21} for details on how $\gamma_j$ is related to the number of crossings and the Zak phases of the isolated bands (if any).

\subsection{Impedance functions in the band gap}
We now briefly recall the concept of impedance function (see \cite{chan-14, lin-zhang-21}), which will be used in the proof of the existence of interface modes in the subsequent subsection.  It is straightforward to see that for each $E$ in the band gap, all the solutions to the equation $(\Lo-E)\psi=0$ with finite $L^2$-norm over the left half-line $(-\infty, 0]$ span a one-dimensional space. Let $\psi_{L, E}$ be one of these solutions. We define the \textbf{impedance function} for the operator $\Lo$ defined over the left half-line to be
$$
\xi_L(E) := \frac{\psi_{L, E}(0)}{ \frac{1}{\mu(0)} \psi_{L, E}'(0)}, \quad \mbox{if} \,\,\psi_{L, E}'(0)\neq 0. 
$$
In the case where $\psi_{L, E}'(0)=0$, $\psi_{L, E}$ has a Neumman boundary condition at $x=0$ and we set formally $\xi_L(E) = \infty$.
Note that $\psi_{L, E}'(0)$ and $\psi_{L, E}(0)$ cannot vanish simultaneously, otherwise $\psi_{L,E}\equiv0$. 
Also, $\xi_L(E)$ defined above is independent of the choice of the solution $\psi_{L, E}$.   
In a similar way, we define the impedance function for the periodic operator $\Lo$ defined on the right half-line $[0, \infty)$ by
$$
\xi_{R}(E) := \frac{\psi_{R, E}(0)}{ \frac{1}{\mu(0)} \psi_{R, E}'(0)}. 
$$
where $\psi_{R, E}$ is a finite $L^2$-norm solution over the right half-line $[0, \infty)$.

We now derive some useful properties of the impedance functions $\xi_{L}(E), \xi_{R}(E) $.

\medskip

\begin{lem} \label{lem-edgemode2}
Let the operator $\Lo$ be of the form (\ref{eq-photonic}). Assume that there is a band gap between the $j$-th and the $(j+1)$-th bands. Then
the following hold for $E \in (E_j^+, E_{j+1}^-)$:
\begin{enumerate}
\item [(i)]
If the Bloch mode at the band gap edge $(k^*,E_j^+)$ has odd-parity for $k^*=0$ or $\pi$, 
then $\xi_{R}(E)$ is strictly decreasing, with $\xi_R(E) \to 0$ as $E \to E_j^+$ and $\xi_R(E) \to -\infty$ as $E \to E_{j+1}^-$;
On the other hand,  $\xi_{L}(E)$ is strictly increasing, with $\xi_L (E)\to 0$ as $E \to E_j^+$ and $\xi_L(E) \to +\infty$ as $E \to E_{j+1}^-$.
\item [(ii)]
If the Bloch mode at band gap edge $(k^*,E_j^+)$ has even-parity, then $\xi_{R}(E)$ is strictly decreasing, with $\xi_R(E) \to +\infty$ as $E \to E_j^+$ and $\xi_R(E) \to 0$ as $E \to E_{j+1}^-$; On the other hand,  $\xi_{L}(E)$ is strictly increasing, with $\xi_L(E) \to -\infty$ as $E \to E_j^+$ and $\xi_L(E) \to 0$ as $E \to E_{j+1}^-$.
\end{enumerate}
\end{lem}

\noindent\textbf{Proof.}  Without loss of generality, we consider only the case $k^*=0$, and 
odd-parity Bloch mode at $(0, E_j^+)$.
The proof for the other cases is similar. It also suffices to consider the function $\xi_R(E)$, since $\xi_L(E)$ can be treated similarly. To further simplify the notations, we assume without loss of generality that $\mu(0)=1$ ($\mu(x)$ is necessarily continuous at $x=0$ due to the inversion symmetry assumption).

Step 1. We first construct a smooth family of real-valued $L^2[0, \infty)$ solutions, denoted by $u_E(x)$, to the equation $(\Lo-E)u=0$ for $E\in (E_j^+, E_{j+1}^-)$. This can be done since the dimension of $L^2[0, \infty)$ solutions to $(\Lo-E)u=0$ is equal to the constant one.  
We refer to \cite{lin-zhang-21} for a concrete construction. Noting that $u_E(0)$ and $u_E'(0)$ cannot be zero simultaneously, we may normalize $u_E$ by requiring that $u_E(0)^2 + u_E'(0)^2 =1$. We also note that as $E$ tends to the band gap edges at $E_j^+$ and at $E_{j+1}^-$, the function $u_E$ tends to the corresponding Bloch modes.

Step 2. We claim that both $u_E(0)$ and $u_E'(0)$ cannot be zero for all $E \in (E_j^+, E_{j+1}^-)$. 
We only prove that $u_E(0)\neq 0$. The claim that $u_E'(0) \neq 0$ can be proved similarly. 
We prove by contradiction. Suppose $u_E(0)=0$ for some $E \in (E_j^+, E_{j+1}^-)$. 
Using the inversion symmetry of the operator $\Lo$, we can check that the function
$\tilde{u}(x)$ defined by $\tilde{u}(x) =u_E(x)$ for $x >0$ and  $\tilde{u}(x) =-u_E(-x)$ for $x <0$ satisfies the equation $(\Lo -E) \tilde{u}=0$ on the whole real line. Moreover $\tilde{u}\in L^2(\mathbf{R})$. Therefore, we see that 
$E$ is a point spectrum of the operator $\Lo$ with eigenfunction 
$\tilde{u}$. This contradicts the fact that $\Lo$ has no point spectrum. 
This completes the proof of the claim. 

Step 3. By the result in Step 2, we can conclude that 
$\xi_R(E) = \frac{u_E(0)}{u'_E(0)}$ is well-defined and is smooth for $E \in (E_j^+, E_{j+1}^-)$. Moreover, $\xi_R(E)$ cannot change signs in $(E_j^+, E_{j+1}^-)$. We now show that $\xi_R(E)$ is strictly decreasing for $E \in (E_j^+, E_{j+1}^-)$. 
Denote 
$ v_E(x) = \frac{\partial u_E(x)}{\partial E}$. By taking the partial derivative with respect to $E$ on both sides of the equation $(\Lo-E)u_E(x)=0$, we obtain
$$
(\Lo -E)v_E(x) = u_E(x). 
$$
Therefore 
$$
\int_0^{\infty} (\Lo-E) v_E(x)\varepsilon(x) u_E(x) dx = \int_{0}^{\infty} \varepsilon(x) u_E(x)^2 dx >0.   $$
Using integration by parts twice and the right decaying property of $u_E(x)$, we see that 
\begin{align*}
\int_0^{\infty} (\Lo-E) v_E(x)\varepsilon(x) u_E(x) dx  &= \int_0^{\infty} (\Lo-E) u_E(x)\varepsilon(x) v_E(x) dx  +  v_E'(0)u_E(0) - v_E(0) u'_E(0) \\
&=  v_E'(0)u_E(0) - v_E(0) u'_E(0). 
\end{align*}
We further obtain 
$$
u'_E(0) \left(v_E'(0) \xi_R(E) -v_E(0)\right)= v_E'(0)u_E(0) - v_E(0) u'_E(0) >0,
$$
where we used the identity
$$
u_E(0) = \xi_R(E) u'_E(0). 
$$
By taking partial derivative respect to $E$ on both sides of the above identity, we obtain
$$
v_E(0) = \xi'_R(E) u'_E(0) + \xi_R(E) v'_E(0).
$$
Therefore
$$
u'_E(0) \left(v_E'(0) \xi_R(E) -v_E(0)\right)
= u'_E(0) \cdot \left(-\xi'_R(E) u'_E(0)\right) = -(u'_E(0) )^2 \xi'_R(E). 
$$
It follows that $\xi'_R(E) < 0$ for all $E \in (E_j^+, E_{j+1}^-)$. This completes the proof of the claim.

Step 4. Finally, using the fact that the Bloch mode at $(k^*=0, E_j^+)$ is odd, we see that 
$u_E(0) \to 0$ as $E \to E_j^+$. Therefore $\xi_R(E) =\frac{u_E(0)}{u'_E(0)}  \to 0$ as $E \to E_j^+$. By the result in Step 3, we conclude that $\xi_R (E)<0$ for all $E \in (E_j^+, E_{j+1}^-)$. On the other hand, 
By Proposition \ref{thm-parity_change}, the Bloch mode at $(k^*=0, E_{j+1}^-)$ is even. Therefore 
$u_E'(0) \to 0$ as $E \to E_{j+1}^-$ and we can conclude that
$\xi_R(E) \to -\infty$ as $E \to E_{j+1}^-$. This completes the proof of the Lemma.

\subsection{Interface modes induced by bulk topological indices}
We consider a photonic system that consists of two semi-infinite periodic structures, one for $x<0$ and one for $x>0$.
The corresponding two periodic differential operators are assumed to be of the form (\ref{eq-photonic}): 
$$
 \Lo_j\psi = -\frac{1}{\varepsilon_j(x)} \dfrac{d}{dx} \left( \frac{1}{\mu_j(x)}\frac{d\psi}{dx}\right), \quad j=1, 2.
$$
The differential operator for the joint structure is given by
\begin{equation}\label{eq-optA}
\tilde \Lo \psi (x) := 
\begin{cases}
 \Lo_1\psi (x) , \quad x <0, \\ 
 \Lo_2\psi (x) , \quad x >0.
 \end{cases}
\end{equation}

We investigate the existence of interface modes for the operator $\tilde \Lo$. Here an interface mode is defined to be a function $\psi$ such that
$$
\psi \in L^2(\R) \quad\mbox{and} \quad (\tilde \Lo -E)\psi  =0
$$
for some real number $E$.
In what follows, we denote the quantities associated with the operator $\Lo_j$ using the superscript $j$ ($j=1, 2$), such as 
the energy level $E_{m}^{(j)}$, the Bloch mode $\varphi_{m,k}^{(j)}$, etc.
Before we proceed, we recall a Lemma that uses impedance functions to prove the existence of an interface mode. Its proof is straightforward. See also \cite{lin-zhang-21}.  

\begin{lem} \label{lem-interface}
Assume that $E$ lies in a common spectral band gap of $\Lo_{1}$ and $\Lo_{2}$, and let $\xi_{L}^{(1)}(E)$ and $\xi_{R}^{(2)}(E)$ be the corresponding impedance functions at the interface $x=0$. Then there exists an interface mode at energy level $E$ for the operator $\tilde \Lo$ if and only if
$$
\xi_{L}^{(1)}(E) =\xi_{R}^{(2)}(E).
$$
\end{lem}

We now are ready to state our main result.

\begin{thm}\label{thm-existence_int_mode}
Assume that the following holds:
\begin{enumerate}
\item[(i)]
The operators $\Lo_1$ and $\Lo_2$ are of the form (\ref{eq-photonic}) and attain a common band gap
$$
I:=( E_{m_1}^{(1),+}, E_{m_1+1}^{(1),-}) \cap (E_{m_2}^{(2),+}, E_{m_2+1}^{(2),-}) \neq \emptyset
$$
for certain positive integers $m_1$ and $m_2$. 
\item[(ii)]
With respect to this common band gap, the bulk topological indices differ, $\gamma_{m_1}^{(1)}\neq \gamma_{m_2}^{(2)}$.
\end{enumerate}
Then there exists a unique interface mode for the operator $\tilde \Lo$ defined in \eqref{eq-optA}. 
\end{thm}

\begin{figure}
    \centering
    \includegraphics[scale=0.7] {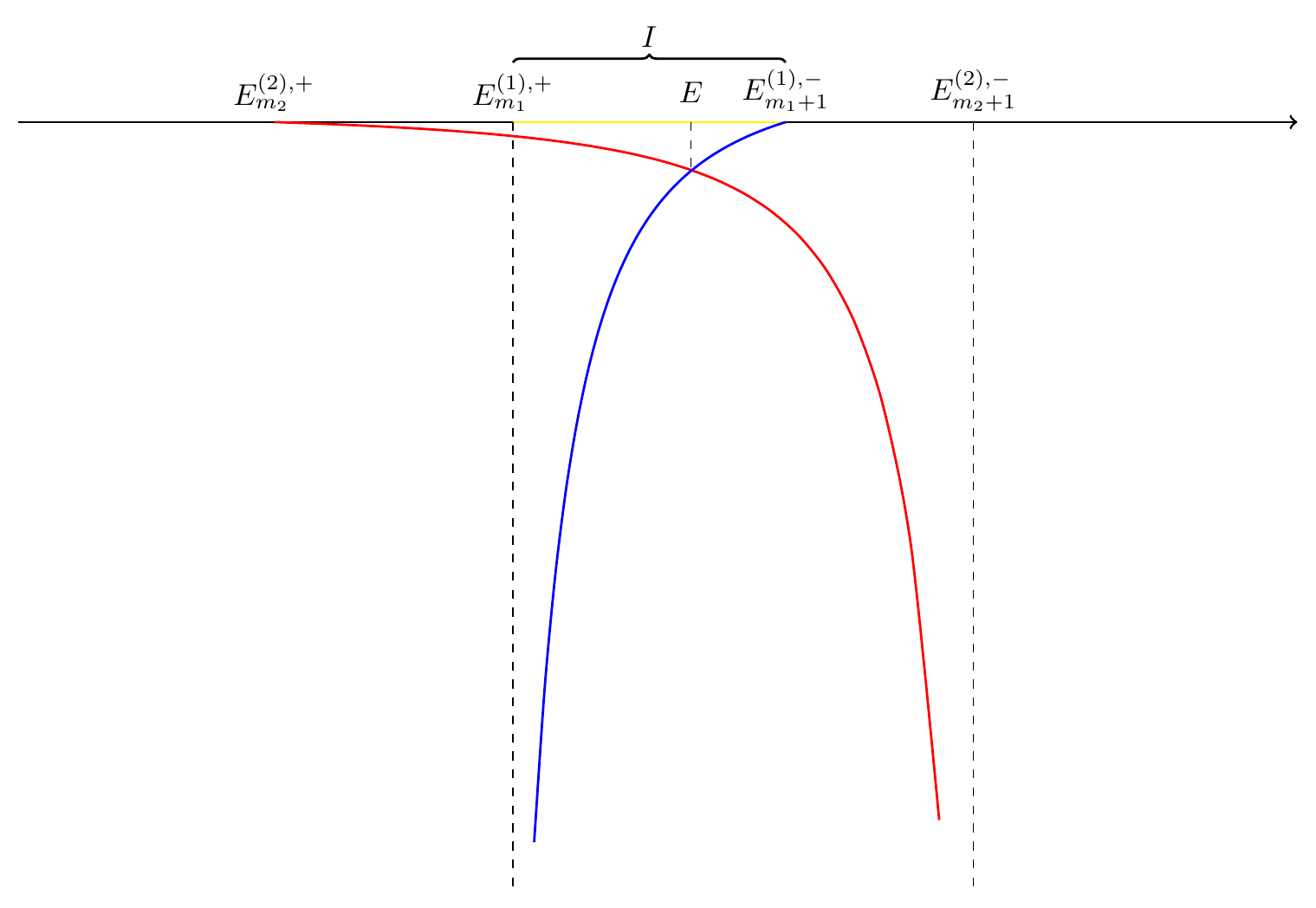}
    \caption{Illustration of proof of Theorem \ref{thm-existence_int_mode}: the red curve is the graph of the impedance function $\xi_R^{(2)}(E)$ defined in the interval $[E_{m_2}^{(2),+}, E_{m_2+1}^{(2),-})$, and the blue curve is that of the impedance function $\xi_L^{(1)}(E)$ defined in the interval
    $(E_{m_1}^{(1),+}, E_{m_1+1}^{(1),-}]$. The two curves intersect at a unique $E\in I=:(E_{m_1}^{(1),+},E_{m_1+1}^{(1),-})$ 
    which gives the interface mode's eigenvalue.}\label{fig:impedance}
\end{figure}

\noindent\textbf{Proof.}  By Lemma \ref{lem-interface}, there is an interface mode of $\tilde{\Lo}$ at energy level $E$ if and only if 
$$
\xi(E):=\xi_{L}^{(1)}(E) - \xi_{R}^{(2)}(E)=0.
$$
Without loss of generality, we consider the case when the common band gap of the operators $\Lo_1$ and $\Lo_2$ is given by $I=(E_{m_1}^{(1),+}, E_{m_1+1}^{(1),-})$.
Moreover,  $\gamma_{m_1}^{(1)}=1$ and $\gamma_{m_2}^{(2)}=-1$ for the two operators.
Then the Bloch mode $\varphi_{m_1,k^*}^{(1)}$ at the band gap edge $(k^*,E_{m_1}^{(1),+})$, where $k^*=0$ or $\pi$, for the operator $\Lo_1$
is even while the Bloch mode $\varphi_{m_2,k^*}^{(2)}$ at the band gap edge $(k^*,E_{m_2}^{(2),+})$ for the operator $\Lo_2$  is odd.
By Lemma \ref{lem-edgemode2},   $\xi_L^{(1)}(E) <0$ and $\xi_L^{(1)}(E) \to -\infty$ as $E \to E_{m_1}^{(1),+}$ and $\xi_L^{(1)}(E) \to 0$ as $E \to E_{m_1+1}^{(1),-}$ respectively. 
On the other hand, $\xi_R^{(2)}(E) <0$ and $\xi_R^{(2)}(E) \to 0$ as $E \to E_{m_2}^{(2),+}$ and $\xi_R^{(2)}(E) \to -\infty$ as $E \to E_{m_2+1}^{(2),-}$ respectively.
Therefore, for $E$ in the common band gap $I$, we see that 
$\xi(E)<0$ for $E$ near $E_{m_1}^{(1),+}$ and $\xi(E)>0$ for $E$ near $E_{m_1+1}^{(1),-}$. Moreover, $\xi(E)$ is strictly increasing since $\xi_L$ is strictly increasing and $\xi_R$ is strictly decreasing. 
It follows that there exists a unique root over the interval $I$ for $\xi(E)=0$.  See Fig.\ \ref{fig:impedance} for an illustration.
\qed \\

As an application of the above theorem, we consider a dislocation model.

\begin{prop}\label{prop:dislocation.mode}
Let $\Lo_1$ be of the form (\ref{eq-photonic}). Assume that the spectrum of $\Lo_1$ has a band gap between the $j$-th and $(j+1)$ band. Further assume that the maximum of the $j$-th spectral band and the minimum of the $(j+1)$-th spectral band are achieved at $k=\pi$. 
Let $\Lo_2$ be the one-half shifted version of $\Lo_1$, in the sense that the corresponding coefficients $\varepsilon_2$ and $\mu_2$ are related to those of $\Lo_1$ by 
$$
\varepsilon_2(x) = \varepsilon_1(x-1/2); \quad \mu_2(x) = \mu_1(x-1/2). 
$$
Then there exists a unique interface mode in the band gap between the $j$-th and $(j+1)$ band for the glued operator $\tilde{\Lo}$ as defined in (\ref{eq-optA}).
\end{prop}

\noindent\textbf{Proof.} 
It is clear that the spectrum of $\Lo_1$ and $\Lo_2$ have the same band structure. Using Theorem \ref{thm-existence_int_mode}, 
we need only to show that the Bloch mode $\varphi_{j, \pi}^{(1)}$ for the operator $\Lo_1$ and the Bloch mode $\varphi_{j, \pi}^{(2)}$ for the operator $\Lo_2$
have different parities. Indeed, up to a constant, the Bloch mode $\varphi_{j, \pi}^{(2)}$ is related to $\varphi_{j, \pi}^{(1)}$ by the following formula 
$$
\varphi_{j, \pi}^{(2)}(x) = \varphi_{j, \pi}^{(1)}(x-1/2).
$$
If $\varphi_{j, \pi}^{(1)}$ is odd, then 
$$
\varphi_{j, \pi}^{(2)}(-x) = \varphi_{j, \pi}^{(1)}(-x-1/2) = -\varphi_{j, \pi}^{(1)}(x+1/2)=\varphi_{j, \pi}^{(1)}(x-1/2) = \varphi_{j, \pi}^{(2)}(x), 
$$
i.e., $\varphi_{j, \pi}^{(2)}$ is even. Similarly, one can show that $\varphi_{j, \pi}^{(2)}$ is odd if $\varphi_{j, \pi}^{(1)}$ is even. This completes the proof of the proposition. 
\qed

\begin{figure}
    \centering
    \begin{tikzpicture}[scale=0.8]
    \begin{axis}[smooth,
xmin = -3.1416, xmax = 3.1416,
    ymin = -2.5, ymax = 2.5]    
        \addplot[blue] {1.25+cos(deg(x))};
        \addplot[blue] {-(1.25+cos(deg(x)))};
                \addplot[dotted] {0};
    \end{axis}
    \filldraw[black] (6.85,2.85) circle (2pt);
    \end{tikzpicture}
    \caption{Spectral dispersion curves for the discrete SSH model with $s=1, t=0.5$, which has an interface zero-energy mode (black dot) according to Prop.\ \ref{prop:SSH}; compare Fig.\ 1 of \cite{ssh-79}. The same dispersion curves are obtained for $s=0.5, t=1$, but the Zak phase is different.
    }
    \label{fig:interface.mode}
\end{figure}
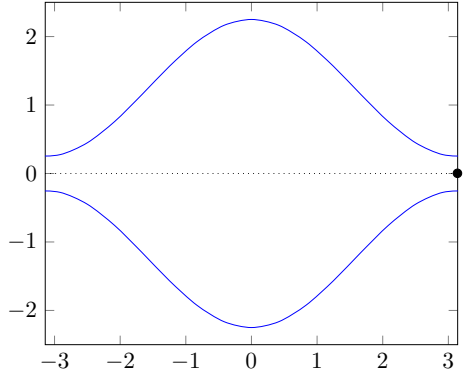

\begin{rem}
Generally, a shift of origin by $1/2$ will change Zak phases by $\pi$ (e.g., \cite{moore-17}), as is apparent from the polarization interpretation of the Zak phase \cite{Vanderbilt-18}; see Lemma \ref{lem:phase.shift} for the same shift in discrete models. Regarding the assumption that the band gap edge occurs at $k=\pi$, see Theorem 2.5 in \cite{lin-zhang-21}, and Fig.\ \ref{fig:interface.mode} for an analogous situation in the SSH lattice model \cite{ssh-79}.
\end{rem}

\begin{rem}
Theorem \ref{thm-existence_int_mode} can be extended to electronic systems modelled by Schr\"{o}dinger operators. More precisely, 
by replacing the operators $\Lo_j$ of the form (\ref{eq-photonic}) with the following ones
$$
\Lo_j = -\frac{d^2}{dx^2} +V_j, \quad j=1,2,
$$
where $V_j$ are real-valued piecewise continuous functions in one dimension that are periodic with period one and are even, the statement of Theorem \ref{thm-existence_int_mode} remains true. This follows from the same arguments. 

\end{rem}


\section{Discrete models}\label{sec:discrete.models}
\subsection{Basic setup and notation}
Throughout, we use the Pauli matrices,
\[
\sigma_1=\begin{pmatrix}0 & 1 \\ 1 & 0\end{pmatrix},\qquad \sigma_2=\begin{pmatrix}0 & -i \\ i & 0\end{pmatrix},\qquad \sigma_3=\begin{pmatrix} 1 & 0 \\ 0 & -1\end{pmatrix},
\]
which satisfy the anticommutation relations 
$
\{\sigma_i,\sigma_j\}\equiv\sigma_i\sigma_j+\sigma_j\sigma_i=2\delta_{ij}.
$
Any real linear combination
\begin{equation}
\boldsymbol{n}\cdot\boldsymbol{\sigma}=n_1\sigma_1+n_2\sigma_2+n_3\sigma_3,\qquad n_1^2+n_2^2+n_3^2=1,\label{eqn:spin.matrix}
\end{equation}
is called a \textbf{spin matrix} with axis along $\boldsymbol{n}=(n_1,n_2,n_3)$.

\medskip

The two-band lattice model Hilbert space is the tensor product $\ell^2(\ZZ)\otimes \CC^2$, with right translation operator denoted $\mathsf{S}\otimes 1$, or simply $\mathsf{S}$. The general $\mathsf{S}$-invariant finite-range self-adjoint Hamiltonian is
\begin{equation}
H=H(A_1,\ldots,A_r,V)=\sum_{i=1}^r \left(\mathsf{S}^i\otimes A_i^*+(\mathsf{S}^*)^i\otimes A_i\right) + 1\otimes V,\label{eqn:bulk.model}
\end{equation}
where $A_i$ is the left-hopping matrix with range $i$, and $V=V^*$ is the on-site potential; they are $2\times 2$ matrices. For convenience, we will often simply write
\[
H=H(A,V)=H(A_1,\ldots,A_r,V).
\]

When $r=1$, we have a \textbf{nearest-neighbour} model. The action of $H=H(A,V)$ on a general $\CC^2$-valued sequence $\psi=(\psi_n)_{n\in\ZZ}$ is then
\[
(H\psi)_n=A^*\psi_{n-1}+A \psi_{n+1}+V\cdot\psi_n,\qquad n\in\ZZ.
\]
Due to $\mathsf{S}$-invariance, we can Fourier transform $H$ into the family of $2\times 2$ Bloch Hamiltonians,
\[
h(k)=A^*e^{ik}+Ae^{-ik}+V,\qquad k\in\mathcal{B}=[-\pi,\pi]/_{-\pi\sim\pi},
\]
where each $h(k)$ acts on the Bloch vector $\psi(k)\in \CC^2$. In the finite-range case, $h$ will be a (matrix-valued) Laurent polynomial in $e^{ik}$. More generally, if $H$ is \textbf{approximately finite-range} in the sense of being approximated in operator norm by Eq.\ \eqref{eqn:bulk.model}, then its Fourier transform $h$ is a continuous $2\times 2$ Hermitian matrix-valued function on $\mathcal{B}$.

The eigenvalues of $h(k)$ vary continuously with $k\in\mathcal{B}$. In total, ${\rm Spec}(H)$ is a closed interval, or a union of two closed intervals, and is purely essential spectrum. In the latter case, the two spectral intervals are separated by a spectral gap $(E_-,E_+)$, and we say that $H$ is \textbf{gapped}. Taking $E\in(E_-,E_+)$, we have $H-E$ being a \emph{Fredholm} operator with empty discrete spectrum. 

Let $\NN$ denote the positive integers. The truncation of $H(A,V)$ to the Hilbert subspace $\ell^2(\NN)\otimes\CC^2$ is denoted $\hat{H}_R=\hat{H}_R(A,V)$. For example, in nearest-neighbour models, we have
\begin{equation}
(\hat{H}_R(A,V)\psi)_n=\begin{cases}
A^*\psi_{n-1}+A \psi_{n+1}+V\psi_n,\qquad\qquad & n\geq 2,\label{eqn:half.space.H}\\
A \psi_{2}+V\psi_1, &n=1.\nonumber
\end{cases}
\end{equation}
Similarly, the truncation of $H(A,V)$ to $\ell^2(-\NN)\otimes\CC^2$ is denoted $\hat{H}_L=\hat{H}_L(A,V)$.

\subsection{Chiral symmetry and index}\label{sec:chiral.index}
Let $\Gamma$ be a grading operator on a Hilbert space, i.e., an operator satisfying $\Gamma=\Gamma^*=\Gamma^{-1}$. It decomposes the Hilbert space into the direct sum of its $+1$ and $-1$ eigenspaces. We will use the symbols $\circ$ and $\bullet$ to represent degrees of freedom from each graded subspace.

Generally, a Hamiltonian operator $H=H^*$ is said to be \textbf{chiral symmetric}, with respect to $\Gamma$, if $H\Gamma=-\Gamma H$ holds. So $H$ has an off-diagonal representation,
\[
H=\begin{pmatrix} 0 & H_{-+}\\ H_{+-} & 0 \end{pmatrix},\qquad H_{-+}=H_{+-}^*,
\]
and comprises terms ``hopping'' between $\circ$ and $\bullet$, but not between $\circ,\circ$ or between $\bullet,\bullet$. A chiral symmetric Hamiltonian has spectrum which is symmetric about 0. Furthermore, if $H$ is Fredholm, then its \textbf{index} is defined to be the usual Fredholm index of $H_{+-}$. In other words, 
\begin{equation}
{\rm Ind}(H):={\rm Ind}(H_{+-})\equiv\dim\ker H_{+-} - \dim\ker H_{-+}\;\in\;\ZZ.\label{eqn:chiral.index}
\end{equation}
So the index of $H$ counts the number of zero-energy modes, with a $+$ sign for $\circ$ and a $-$ sign for $\bullet$. The index problem is to find a formula for ${\rm Ind}(H)$ in terms of a topological invariant associated to $H$.

\subsubsection*{Chiral symmetry: Index formula via winding number}
Returning to lattice models on $\ell^2(\ZZ)\otimes\CC^2$, a \textbf{sublattice operator} is a grading operator of the form $\Gamma=1\otimes\gamma$. The $\CC^2$ at each unit cell is split into the $\pm$ eigenspaces of the $2\times 2$ matrix $\gamma$. So we have a $\circ$ sublattice, and a $\bullet$ sublattice,
\[
\cdots |\,\circ\;\;\bullet\;|\,\circ\;\;\bullet\,|\;\circ\;\;\bullet\;|\cdots\nonumber\\
\]
For $H(A,V)$ to be chiral symmetric, the matrices $A,V$ have to anticommute with $\gamma$, thus they are off-diagonal (in a basis where $\gamma=\sigma_3$). Then the Fourier transform of $H(A,V)$ has the form
\begin{equation}
h(k)=\begin{pmatrix} 0 & h_{-+}(k)\\ h_{+-}(k) & 0\end{pmatrix}=\begin{pmatrix} 0 & \overline{h_{+-}(k)}\\ h_{+-}(k) & 0\end{pmatrix},\qquad k\in\mathcal{B}.\label{eqn:chiral.symbol}
\end{equation}
We call $h_{+-}:\mathcal{B}\to \CC$ of Eq.\ \eqref{eqn:chiral.symbol} the \textbf{symbol} function of $H(A,V)$. Since $h(k)^2=|h_{+-}(k)|^2\,\mathbf{1}_2$, the eigenvalues of $h(k)$ are $\pm|h_{+-}(k)|$. Therefore $H(A,V)$ has a spectral gap around $0$ if and only if its symbol function $h_{+-}$ is nowhere-vanishing. The homotopy class of $h_{+-}:\mathcal{B}\to \CC^*$, i.e., its winding number, is the \textbf{bulk topological index} for $H$.

Note that $\ell^2(\NN)\subset \ell^2(\ZZ)$ is identified with the Hardy subspace $H^2(\mathcal{B})\subset L^2(\mathcal{B})$ via Fourier transform. So $\hat{H}_{R,+-}:\ell^2(\NN)\to\ell^2(\NN)$ is identified with the compression of $H_{+-}:\ell^2(\ZZ)\to\ell^2(\ZZ)$ to the Hardy space. In other words, $\hat{H}_{R,+-}$ is identified with the classical Toeplitz operator on $H^2(\mathcal{B})\cong \ell^2(\NN)$ with continuous symbol function $h_{+-}$. The $C^*$-algebra $\mathcal{T}$ of Toeplitz operators on $\ell^2(\NN)$ lies in a short exact sequence of $C^*$-algebras (see (4.9) of \cite{arv-02}),
\begin{equation}
0\longrightarrow \mathcal{K}\longrightarrow\mathcal{T}\overset{\rm symbol}{\longrightarrow}C(\mathcal{B})\longrightarrow 0,\label{eqn:Toeplitz.SES}
\end{equation}
where $\mathcal{K}$ denotes the compact operators on $\ell^2(\NN)$. Similarly for Toeplitz operators acting on $\ell^2(\NN)\otimes\CC^2$, such as $\hat{H}_R(A,V)$. Writing $M_2(\CC)$ for the algebra of $2\times 2$ matrices, we have $H(A,V)\overset{\rm Fourier}{\cong} h\in C(\mathcal{B})\otimes M_2(\CC)$, while $\hat{H}_R(A,V)\in\mathcal{T}\otimes M_2(\CC)$. Then Eq.\ \eqref{eqn:Toeplitz.SES} implies that the spectrum of $H(A,V)$ is precisely the spectrum of $\hat{H}_R(A,V)$ modulo the compact operators.

Therefore, if $H(A,V)$ is chiral symmetric and has a spectral gap around $0$, then the essential spectrum of $\hat{H}_R(A,V)$ is likewise gapped around $0$. In this situation, $\hat{H}_R(A,V)$ is Fredholm, and we may ask for its index in the sense of Eq.\ \eqref{eqn:chiral.index}, ${\rm Ind}\,\hat{H}_R(A,V)\equiv {\rm Ind}\,\hat{H}_{R,+-}$. The Toeplitz index theorem (see, e.g., Theorem 4.4.3 of \cite{arv-02} for a proof) says that
\begin{equation}
{\rm Ind}(\hat{H}_R(A,V))=-{\rm Wind}(h_{+-})=-{\rm Ind}(\hat{H}_L(A,V)).\label{eqn:winding.index.signed}
\end{equation}
Actually, even more is true: either $\ker \hat{H}_{R,+-}=0$ or $\ker \hat{H}_{R,-+}=0$, see \cite{coburn-66} and Theorem 4.5.4 of \cite{arv-02}. So the (unsigned) kernel dimension of $\hat{H}_R(A,V)$ is given by the absolute value of its index. Similarly for $\hat{H}_L(A,V)$. To summarize, we have proved:
\begin{thm}\label{thm:BEC.chiral}
Let $H(A,V)$ be an approximately finite-range Hamiltonian which is chiral symmetric and gapped (two-band model). Then 
\begin{equation*}
\dim \ker \hat{H}_R =|{\rm Wind}(h_{+-})|=\dim \ker \hat{H}_L.\label{eqn:winding.index.unsigned}
\end{equation*}
\end{thm}

\medskip

In nearest-neighbour models, one just has a second-order difference equation, Eq.\ \eqref{eqn:half.space.H}, and Theorem \ref{thm:BEC.chiral} can be shown by direct algebraic means. In fact, one obtains a supplementary statement on the in-gap eigenvalues, proved in Theorems 1a--b of \cite{mong-11}, see also Theorem 10 of \cite{Shapiro}:
\begin{prop}\label{prop:NN.sole.edge.state}
Let $H(A,V)$ be a chiral symmetric nearest-neighbour Hamiltonian with a spectral gap (around $0$). Then $\hat{H}_R(A,V)$ does not have non-zero in-gap eigenvalues. 
\end{prop}
Without the nearest-neighbour assumption, in-gap eigenvalues of $\hat{H}_R(A,V)$ can appear in $\pm E$ pairs. Although Theorem \ref{thm:BEC.chiral} still relates the zero-energy modes with non-trivial winding numbers, the spectral gap between such zero-energy modes and the rest of the spectrum will generally be much more narrow than the bulk spectral gap.

\subsection{$\mathsf{P}$ and $\mathsf{T}$ symmetry, and associated sublattice operator}\label{sec:associated.sublattice}
\paragraph*{$\mathsf{T}$-symmetry}
$H(A,V)$ is said to be \textbf{time-reversal symmetric}, or \textbf{$\mathsf{T}$-symmetric}, if it commutes with the operation $\mathsf{T}$ of complex-conjugation; equivalently, the matrices $A,V$ are real-valued.

\paragraph*{$\mathsf{P}$-symmetry}
The operator $\mathsf{R}$ of inverting position labels $n\leftrightarrow -n$ swaps $\mathsf{S}$ for $\mathsf{S}^*$, so it
effects
\begin{equation*}
\mathsf{R}H(A,V)\mathsf{R}=\sum_{i=1}^r \left((\mathsf{S}^*)^i\otimes A_i^*+\mathsf{S}^i\otimes A_i\right)+1\otimes V=H(A^*,V).
\label{eqn:position.label.reflection}
\end{equation*}
For general reasons, it is more appropriate to use an inversion operator of the form
\[
\mathsf{P}=\mathsf{R}\otimes Q,\qquad (\mathsf{P}\psi)_n:=Q\psi_{-n},\qquad n\in\ZZ,
\]
where $Q$ is some unitary $2\times 2$ unitary matrix having both $+1$ and $-1$ eigenvalues. Such a $Q$ necessarily has the form of a spin matrix, Eq.\ \eqref{eqn:spin.matrix}, and any choice of $Q$ is unitarily related to another by an ${\rm SU}(2)$ spin rotation. Since
\[
\mathsf{P}H(A,V)\mathsf{P}=\sum_{i=1}^r\left((\mathsf{S}^*)^i\otimes QA_i^*Q+\mathsf{S}^i\otimes QA_iQ\right)+1\otimes QVQ=H(QA^*Q,QVQ),
\]
$H(A,V)$ is \textbf{$\mathsf{P}$-symmetric} if and only if $(A_1,\ldots,A_r,V)$ satisfy
\begin{equation}
QA_i^*=A_iQ,\qquad QV=VQ.\label{eqn:hopping.symmetry}
\end{equation}

\paragraph*{Associated sublattice operator $\Gamma$}
Where simultaneously present, $\mathsf{T}$ and $\mathsf{P}$ are assumed to commute. In particular, at $n=0$, this forces $Q$ to be a spin matrix with real entries. Therefore $Q=n_1\sigma_1+n_3\sigma_3$ is constrained to have spin axis lying in the $1$-$3$ plane.
Define the associated sublattice operator to be $\Gamma=1\otimes\gamma$, where $\gamma$ is the spin operator with axis in the $1$-$3$ plane but perpendicular to that of $Q$,
\[
\gamma=-n_3\sigma_1+n_1\sigma_3.
\]
Up to a sign, this sublattice operator is \emph{uniquely} characterised by the conditions
\[
\Gamma=\Gamma^*=\Gamma^{-1},\qquad \Gamma \mathsf{P}=-\mathsf{P}\Gamma,\qquad \Gamma\mathsf{T}=\mathsf{T}\Gamma,
\]
This is because: (i) $\gamma$ has real entries, so it has the form $m_1\sigma_1+m_3\sigma_3$, and (ii) anticommutativity, $\{\mathsf{P},\Gamma\}=0$ thus $\{Q,\gamma\}=0$, forces $(m_1,0,m_3)\perp(n_1,0,n_3)$.

It is instructive to summarize the above discussion pictorially. The operator $\mathsf{P}$ does not only map the unit cell $n$ to the unit cell $-n$, it also exchanges the two sublattices,
\begin{align*}
\cdots &|\,\circ\;\;\bullet\;|\,\circ\;\;\bullet\,|\;\circ\;\;\bullet\;|\cdots\nonumber\\
\cdots &|\,\bullet\;\;\circ\;|\,\bullet\;\;\circ\;|\,\bullet\;\;\circ\;|\cdots\label{eqn:sublattice.arrangement}
\end{align*}
Subsequently, it will be convenient to work in a basis for $\CC^2$ such that
\[
Q=\sigma_1,\qquad \gamma=\sigma_3,\qquad
\circ\sim\binom{1}{0},\;\bullet\sim\binom{0}{1}. 
\]

\begin{rem}\label{rem:reflection.and.sublattice}
The roles of $Q$ and $\gamma$ are interchangeable. That is, if a real sublattice operator $\Gamma=1\otimes\gamma$ is given, then there is a canonical real reflection matrix $Q$ (up to a sign) such that $\mathsf{P}=\mathsf{R}\otimes Q$ exchanges the sublattices.
\end{rem}

\subsection{Topological invariants for inversion and/or chiral symmetric Hamiltonians}\label{sec:relation.between.invariants}

For a gapped $H(A,V)$, the lower energy band is a Hermitian line bundle $\mathcal{E}_-$ over the Brillouin zone $\mathcal{B}$. Specifically, the complex line $\mathcal{E}_{-,k}$ at $k\in\mathcal{B}$ is the negative eigenspace of the Bloch Hamiltonian $h(k)$. Given a connection on $\mathcal{E}_-$, one acquires a ${\rm U}(1)$-valued holonomy when parallel transporting a vector in $\mathcal{E}_-$ around $\mathcal{B}$. In the physics literature, this phase is often referred to as a \emph{Zak phase}, or \emph{Berry phase}. It is customary to call this the Zak phase of $H(A,V)$, and to refer to the argument $\theta$ (mod $2\pi$) in the phase $e^{i\theta}$.

In the context of Bloch electrons, the connection/parallel transport on $\mathcal{E}_-$ is not canonically given, but rather depends on a choice of origin, see \cite{moore-17}. In a lattice model, one usually forgets this subtlety, since an origin is implicitly given by specifying the $n=0$ unit cell (containing $N$ degrees of freedom). Then the Fourier transformed Hilbert space, $L^2(\mathcal{B};\CC^N)$, comprises square-integrable sections $\psi:k\mapsto \psi(k)$ of a trivialized bundle $\mathcal{E}=\mathcal{B}\times\CC^N$, and the subbundle $\mathcal{E}_-$ inherits a Grassmann--Berry connection $\mathcal{A}$, typically represented as $\mathcal{A}(k)\,dk=i\langle \psi(k)|\partial_k \psi(k)\rangle\,dk$ with the $\psi(k)\in\mathcal{E}_{-,k}\subset \CC^N$ smoothly chosen and normalized. Here $\partial_k$ makes sense as the trivial connection on the trivialized $\mathcal{E}$. Notwithstanding this, a different choice of unit cell does change the implied origin, and therefore the connection and its Zak phase. We will encounter this ambiguity in Section \ref{sec:unit.cell.convention}.

\subsubsection*{Chiral symmetry: winding number versus Zak phase}
The winding number of $h_{+-}$ is related to $\mathcal{E}_-$ and its Zak phase as follows. Write
\begin{equation}
\frac{h(k)}{|h(k)|}=\begin{pmatrix} 0 & \overline{z(k)}\\ z(k) & 0 \end{pmatrix},\qquad z(k):=\frac{h_{+-}(k)}{|h_{+-}(k)|}\in{\rm U}(1).\label{eqn:normalized.symbol}
\end{equation}
The negative-energy eigenspace of $h(k)$, namely $\mathcal{E}_{-,k}$, is precisely the $-1$ eigenspace of $\frac{h(k)}{|h(k)|}$. This eigenspace is easily checked to be
\[
\mathcal{E}_{-,k}={\rm span}\left\{\frac{1}{\sqrt{2}}\binom{1}{-z(k)}\right\},\qquad k\in\mathcal{B}.
\]
So on the negative-energy eigenbundle $\mathcal{E}_-$, the Berry connection 1-form is
\[
\mathcal{A}=\frac{i}{2}\begin{pmatrix} 1 & -\overline{z}\end{pmatrix}\cdot\partial_k\binom{1}{-z}\,dk=\frac{i}{2}\,\overline{z}\,\frac{dz}{dk}\,dk=\frac{i}{2}z^{-1}dz,
\]
which integrates over $\mathcal{B}$ to
\[
\int_\mathcal{B} \mathcal{A}=\pi\cdot\frac{i}{2\pi}\int_\mathcal{B}z^{-1}dz=-\pi\cdot {\rm Wind}(z).
\]
Modulo $2\pi$, the Zak phase is therefore equal to $\pi$ times of the winding number of the map $z:\mathcal{B}\to{\rm U}(1)$, or equivalently, that of the symbol $h_{+-}:\mathcal{B}\to \CC^*$. To summarize,
\begin{equation}
{\rm Zak}(H(A,V))=\pi\Big({\rm Wind}(h_{+-})\;\;{\rm mod}\; 2\Big).\qquad\;\; \mathrm{(chiral\;symmetry\;present)}\label{eqn:winding.Zak}
\end{equation}

\subsubsection*{$\mathsf{P}$ or $\mathsf{PT}$ symmetry and quantization of Zak phase}
On the Fourier transform, $\mathsf{P}$ acts on $\psi=\psi(k)$ as
\begin{equation*}
(\mathsf{P}\psi)(k)=Q\psi(-k)\label{eqn:P.action}
\end{equation*}
It is well-known that one-dimensional inversion-symmetric gapped Hamiltonians have Zak phases quantized to values $0$ or $\pi$, see \cite{Zak-89}.

Let us explain why the same quantization occurs in the presence of $\mathsf{PT}$ symmetry. With $(\mathsf{T}\psi)(k)=\overline{\psi(-k)}$, we have
\begin{equation}
(\mathsf{PT}\psi)(k)=(\mathsf{TP}\psi)(k)=\overline{Q\psi(k)},\qquad\qquad\qquad\qquad k\in\mathcal{B}.\label{eqn:PT.action}
\end{equation}
Thus $\mathsf{PT}$ defines a \emph{real structure} (generalized complex conjugation) on each space of $k$-quasiperiodic Bloch modes, $k\in\mathcal{B}$. Suppose $H(A,V)$ is gapped and commutes with $\mathsf{PT}$ (but not necessarily with $\mathsf{P}$ and $\mathsf{T}$ separately). Then we can ask for the choice of Bloch eigenvector $\psi(k)$ in each $\mathcal{E}_{-,k}$ to be real with respect to $\mathsf{PT}$, in which case the phase freedom is reduced from ${\rm U}(1)$ to ${\rm O}(1)=\{\pm 1\}$. Thus, the space of $\mathsf{PT}$-real Bloch eigenvectors in $\mathcal{E}_-$ forms a principal ${\rm O}(1)$-bundle over $\mathcal{B}$. There are two possibilities --- the trivial bundle and the M\"{o}bius bundle, distinguished by the ${\rm O}(1)$-valued holonomy ($0,\pi$-valued Zak phase). A $\pi$ Zak phase means that there is no globally continuous choice of $\mathsf{PT}$-invariant eigenvectors for $\mathcal{E}_-$ --- one inevitably acquires a $-1$ mismatch after going around $\mathcal{B}$. 

We stress that the quantization of Zak phase has nothing, a priori, to do with chiral symmetry. Nevertheless, if $H$ is chiral symmetric, then $h$ has the form in Eq.\ \eqref{eqn:normalized.symbol} and it is easily seen to commute with the $\mathsf{PT}$ action (Remark \ref{rem:reflection.and.sublattice}) given by Eq.\ \eqref{eqn:PT.action}. So chiral symmetry implies $\mathsf{PT}$-symmetry. But unlike the winding number, the Zak phase remains invariant even when chiral symmetry is broken, as long as $\mathsf{PT}$-symmetry is retained.

\begin{rem}
As we will be concerned with Hamiltonians which are separately $\mathsf{P}$ and $\mathsf{T}$ symmetric, we point out that the $\mathsf{P}$ symmetry implies that the quantized Zak phase is equivalently determined by the product-of-parities at $k=0$ and $k=\pi$, see \cite{HPB-11}.
\end{rem}

\subsection{Inversion symmetric bulk-edge correspondence}
\subsubsection*{Strictly nearest-neighbour coupling}
Let $H(A,V)$ be $\mathsf{P}$ and $\mathsf{T}$ symmetric. So there is an associated sublattice operator $\Gamma=1\otimes\gamma$, with $\{Q,\gamma\}=0$, according to Section \ref{sec:associated.sublattice}.  
Let us reexamine the sublattice picture,
\[
\cdots |\,\circ\;\;\bullet\;|\,\circ\;\;\bullet\;|\;\circ\;\;\bullet\;|\cdots\nonumber\\
\]
Notice that the nearest-neighbours of $\circ$ are always $\bullet$. So a hopping term between a pair of $\circ$ belonging to adjacent unit cells is actually a \emph{next}-nearest-neighbour coupling. It is natural to require that the dominant terms in $H(A,V)$ are the \emph{strict} nearest-neighbour couplings between adjacent $\bullet,\circ$, together with the on-site potential.

Therefore, we define the \textbf{strictly nearest-neighbour} part of $H(A,V)$, denoted $H^{\rm SNN}(A,V)$, to be given by the on-site potential together with the terms coupling adjacent $\bullet\leftarrow\circ$ and $\circ\to\bullet$. Explicitly, using a basis where $\gamma=\sigma_3, Q=\sigma_1$, we have
\begin{equation}
H^{\rm SNN}(A,V)=\mathsf{S}^*\otimes\begin{pmatrix} 0 & 0 \\ t & 0\end{pmatrix}+\mathsf{S}\otimes\begin{pmatrix} 0 & t \\ 0 & 0 \end{pmatrix} + 1\otimes \begin{pmatrix} v_0 & s \\ s & v_0\end{pmatrix},\label{eqn:SSN.Hamiltonian}
\end{equation}
for some $v_0,s,t\in\RR$ determined by $(A,V)$. By construction, $H^{\rm SNN}(A,V)$ is chiral symmetric up to the overall scalar $v_0$. Observe that $H^{\rm SNN}(A,V)$ is $\mathsf{P}$ and $\mathsf{T}$ symmetric, and therefore, so is the remainder
\[
H^{\rm far}(A,V):=H(A,V)-H^{\rm SNN}(A,V).
\]

With these definitions, we can state and prove the following bulk-edge correspondence:
\begin{thm}\label{thm:PT.BEC}
Let $H(A,V)$ be an approximately finite-range, $\mathsf{P}$ and $\mathsf{T}$ symmetric Hamiltonian. We assume that its strictly nearest-neighbour part, $H^{\rm SNN}(A,V)$, has a spectral gap of size $2\Delta>0$ around some $v_0\in\RR$, and that the remainder $H^{\rm far}(A,V)$ has norm smaller than $\frac{\Delta}{2}$. If the Zak phase of $H(A,V)$ is $\pi$ (resp.\ $0$), then $\hat{H}_R(A,V)$ has one (resp.\ no) in-gap eigenvalue inside the interval \mbox{$(v_0-\frac{\Delta}{2}, v_0+\frac{\Delta}{2})$}.
\end{thm}
\noindent\textbf{Proof.}
As discussed above, $H^{\rm SNN}(A,V)-v_0$ is chiral symmetric, and it has a spectral gap $(-\Delta,\Delta)$ by assumption. Its symbol function is a polynomial in $e^{ik}$ of degree at most $\pm 1$, thus the winding number has magnitude at most $1$. Theorem \ref{thm:BEC.chiral}, together with Eq.\ \eqref{eqn:winding.Zak}, says that $\hat{H}_R^{\rm SNN}(A,V)$ has a (mid-gap) eigenvalue $v_0$ precisely when the Zak phase of $H^{\rm SNN}(A,V)$ is $\pi$. Furthermore, this will be the only eigenvalue in the spectral gap of $H^{\rm SNN}(A,V)$, by Prop.\ \ref{prop:NN.sole.edge.state}, so it will be isolated from the rest of the spectrum of $H^{\rm SNN}(A,V)$ by a distance $\Delta$. If the Zak phase of $H^{\rm SNN}(A,V)$ is $0$, there are no in-gap eigenvalues at all.

Now restore the remainder term $H^{\rm far}(A,V)$. Since its norm is assumed to be smaller than $\frac{\Delta}{2}$, the total operator $H(A,V)$ still has at least $(v_0-\frac{\Delta}{2},v_0+\frac{\Delta}{2})$ as a spectral gap. Therefore the Zak phase of $H(A,V)$ remains well-defined, and coincides with that of $H^{\rm SNN}(A,V)$. Suppose this Zak phase is $\pi$, so we know that $\hat{H}_R^{\rm SNN}(A,V)$ has an in-gap eigenvalue with isolation distance $\Delta$. The half-space Hamiltonian has remainder term $\hat{H}_R^{\rm far}(A,V)$ with norm smaller than $\frac{\Delta}{2}$ (compression to a Toeplitz operator preserves the norm, see Theorem 4.2.4 of \cite{arv-02}). So by spectral perturbation theory, see \S 4.V.3 of \cite{kato-80}, the total half-space operator $\hat{H}_R(A,V)=\hat{H}_R^{\rm SNN}(A,V)+\hat{H}_R^{\rm far}(A,V)$ still has one eigenvalue in the interval $(v_0-\frac{\Delta}{2},v_0+\frac{\Delta}{2})$. Similarly, if the Zak phase is 0, then $(v_0-\frac{\Delta}{2},v_0+\frac{\Delta}{2})$ remains a spectral gap for $\hat{H}_R(A,V)$.
\qed

\subsection{Effect of unit cell convention on Zak phase}\label{sec:unit.cell.convention}
Imagine that the degrees of freedom in the unit cells are embedded in the real line as follows,
\begin{equation*}
\cdots\;\bullet\;\;|\;\;\;\circ\,\bullet\;\;\;\,|\;\;\;\circ\,\bullet\;\;\;\,|\;\;\circ\;\cdots\label{eqn:convention.2}
\end{equation*}
Now shift the unit cell convention by half a unit cell to the right,
\begin{equation*}
\cdots|\bullet\;\;\;\;\,\;\;\circ\,|\bullet\;\;\;\;\,\;\;\circ\,|\bullet\;\;\;\;\,\;\;\circ\,|\cdots\label{eqn:convention.1}
\end{equation*}
There is no physical effect, of course. However, the position labels for the $\circ$ sublattice get shifted by 1, while those for the $\bullet$ sublattice remain unchanged. Thus, we need to apply the unitary transformation $U={\rm diag}({\rm shift}, {\rm id})$ to change conventions.

\begin{lem}\label{lem:phase.shift}
Let $H(A,V)$ be a $\mathsf{P}$ and $\mathsf{T}$ symmetric gapped Hamiltonian, so that it has a $0$ or $\pi$-valued Zak phase with respect to a given unit cell convention. Upon switching to the half unit-cell shifted convention, the Zak phase is shifted by $\pi$.
\end{lem}
\noindent\textbf{Proof.}
As explained in Section \ref{sec:relation.between.invariants}, the Zak phase may be computed as the product-of-parities of the lower energy Bloch modes at $k=0$ and $k=\pi$. Let us write out these Bloch modes explicitly. The matrix $Q=\sigma_1$ has $\pm$ eigenvalues with respective eigenvector $\binom{\pm 1}{1}$. So the periodic Bloch modes (i.e.\ $k=0$) with even/odd parity are
\begin{equation*}
    (\psi^{0,e})_n=\binom{1}{1},\qquad (\psi^{0,o})_n=\binom{-1}{1},\qquad n\in\ZZ.
\end{equation*}
One of these is the lower energy mode for $H(A,V)$ at $k=0$. Similarly, the antiperiodic ($k=\pi$) Bloch modes with even/odd parity are, respectively,
\begin{equation*}
    (\psi^{\pi,e})_n=(-1)^n\binom{1}{1},\qquad (\psi^{\pi,o})_n=(-1)^n\binom{-1}{1},\qquad n\in\ZZ.
\end{equation*}
One of these is the lower energy mode for $H(A,V)$ at $k=\pi$. From the above expressions, $U={\rm diag}({\rm shift},{\rm id})$ acts as the identity on $\psi^{0,e}$ and $\psi^{0,o}$, whereas it exchanges $\psi^{\pi,e}$ with $\psi^{\pi,o}$. Therefore, after applying $U$, the parity of the lower energy mode of $H(A,V)$ at $k=\pi$ is changed, resulting in a $\pi$-shifted Zak phase.
\qed

Lemma \ref{lem:phase.shift} shows that the bulk Zak phase invariant has no observable meaning without reference to a unit cell convention. This subtlety also arises for the winding numbers of chiral symmetric Hamiltonians, and was highlighted in \cite{thiang-15}; see also \cite{SW-21} for a related discussion.
In our bulk-boundary correspondence, Theorem \ref{thm:PT.BEC}, the boundary termination designates the unit cell convention. 

\medskip


\subsection{Interface modes in SSH discrete models}\label{sec:dislocation.model}
We introduce the Hilbert space
\begin{equation}
\ell^2(-\NN;\CC^2)\oplus\CC\oplus\ell^2(\NN;\CC^2),\label{eqn:defect.Hilbert.space}
\end{equation}
corresponding to the following partition,
\begin{alignat}{7}
\cdots\;\; \circ\;\;\,&|\bullet\;\;\,\circ\, &&|\bullet\;\;\,\circ\, &&|\;\;\bullet\;\; &&|\circ\;\;\,\bullet\, &&|\circ\;\;\,\bullet\, &&|\circ\;\;\,\bullet\, &&|\cdots \label{eqn:dislocated.lattice}\\
\cdots -3\;\;\;&|-2 &&|-1 &&|\;\;0 &&|\;\;\;\;1 &&|\;\;\;\;2 &&|\;\;\;\;3 &&|\cdots\nonumber
\end{alignat}
and study \emph{interface Hamiltonians} $H_{\rm int}=H_{\rm int}(A_L,V_L,A_R,V_R;B_L,B_R,W)$, defined as
\begin{equation}
(H_{\rm int}\psi)_n=\begin{cases}
A_R^*\psi_{n-1}+A_R \psi_{n+1}+V_R\psi_n,\qquad\qquad &n\geq 2,\\
B_R^*\psi_0+ A_R \psi_{2}+V_R\psi_1, &n=1,\\
B_L^*\psi_{-1}+B_R\psi_1+W\psi_0, &n=0,\\
A_L^*\psi_{-2}+B_L\psi_0+V_L\psi_{-1}, &n=-1,\\
A_L^*\psi_{n-1}+A_L\psi_{n+1}+V_L\psi_n, &n\leq -2.
\end{cases}\label{eqn:interface.H}
\end{equation}
Here $B_L^*,B_R$ are $1\times 2$ matrices hopping from $n=-1$ to $n=0$, and $n=+1$ to $n=0$ respectively, while $W\in\RR$ is the on-site potential at $n=0$. Put simply, once $n\geq 2$, we have $H(A_R,V_R)$, and once $n\leq -2$, we have $H(A_L,V_L)$. The interface region covers $n=-1,0,+1$, and involves also the hopping terms $B_L,B_R$ and defect potential $W$. The nearest-neighbour interface model of Eq.\ \eqref{eqn:interface.H} is easily generalized to (approximately) finite-range interface Hamiltonians, by replacing $A_L, A_R$ with a sequence of hopping matrices, and $B_L,B_R, W$ with a finitely-supported (thus compact) interface term.

The bulk parameters $(A_L,V_L)$ and $(A_R,V_R)$ are each assumed to satisfy the $\mathsf{P}$-symmetry condition, Eq.\ \eqref{eqn:hopping.symmetry}, and $\mathsf{T}$-symmetry, while the interface terms $(B_L,B_R,W)$ are arbitrary. So there is an associated sublattice operator on the Hilbert space \eqref{eqn:defect.Hilbert.space}, as indicated by the $\circ,\bullet$ symbols in Eq.\ \eqref{eqn:dislocated.lattice}.

\begin{example}\label{ex:simplest.interface}
If $V_L=V_R=\sigma_1$ and $A_L,A_R,B_L,B_R,W=0$, then we have, pictorially,
\[
\cdots|\;\bullet=\circ\;|\;\bullet=\circ\;|\;\bullet\;|\;\circ=\bullet\;|\;\circ=\bullet\;|\cdots
\]
Neglecting $n=0$, the operator $H_{\rm int}$ is just an infinite direct sum of $\sigma_1$, and the bulk spectrum is obviously $\{-1,+1\}$. However, the interface degree of freedom is now a zero-energy mode.
\end{example}

\subsubsection*{Su--Schrieffer--Heeger model}
The classic polymer interface SSH model \cite{ssh-79} is a strictly nearest-neighbour model, pictorially represented as
\begin{alignat}{7}
\cdots\;&\circ\; &-&\;\bullet=\circ\; &-&\;\bullet=\circ\; &-&\;\;\bullet\;\; &-&\;\circ=\bullet\; &-&\;\circ=\bullet\; &-&\circ\cdots\label{eqn:SSH.lattice}\\
\cdots& &|\;&\;\; -2 &|\;&\;\; -1  &|\;&\;\,0 &|\;&\;\;\;\;\; 1 &|\;& \;\;\;\;\;2 &|\;&\;\;\;\;\cdots
\nonumber
\end{alignat}
The intracell hopping term, indicated by $=$, is $V_L=V_R=V=s\sigma_1$ for some amplitude $s\in\RR$. On the right side, the intercell left-hopping term $\bullet\leftarrow\circ$ is
\[
A_R=A=\begin{pmatrix} 0 & 0 \\ t & 0 \end{pmatrix},\qquad t\in\RR.
\]
On the left side, the intercell left-hopping term is $\circ\leftarrow\bullet$, but note that $\bullet\sim\binom{1}{0}$ and $\circ\sim\binom{0}{1}$ on this side, so we still have $A_L=A$. The bulk Hamiltonians $H(A_L,V_L)=H(A,V)=H(A_R,V_R)$ are precisely of the $\mathsf{P}$ and $\mathsf{T}$ symmetric, strictly nearest-neighbour form considered in Eq. \eqref{eqn:SSN.Hamiltonian}, with no on-site term $v_0$. 

The interface terms are
\begin{equation*}
B_R=\begin{pmatrix} u_R & 0 \end{pmatrix},\;\; B_L^*=\begin{pmatrix} 0 & u_L \end{pmatrix},\qquad W=0,\qquad u_L,u_R\in\RR.\label{eqn:SSH.interface.terms}
\end{equation*}
Importantly, the overall \textbf{SSH model Hamiltonian}, 
\begin{equation}
H_{\rm SSH}=H_{\rm int}(A,V,A,V; B_L,B_R,0),\label{eqn:SSH.model.Hamiltonian}
\end{equation}
is chiral symmetric. Example \ref{ex:simplest.interface} is an SSH model Hamiltonian with $s=1, t=0$ and trivial interface terms.

\medskip
To understand the spectrum of $H_{\rm SSH}$, let us first turn off the interface terms, so that we just have a direct sum decomposition
\begin{equation*}
H_{\rm int}(A,V,A,V;0,0,0)=\hat{H}_L(A,V)\oplus 0 \oplus \hat{H}_R(A,V).
\end{equation*}
The bulk Hamiltonian, $H(A,V)$, has nowhere-vanishing symbol function  $h_{+-}(k)=s+te^{-ik}$ (thus $H(A,V)$ is gapped), if and only if $|s|\neq |t|$. When $|s|>|t|$, the winding number is $0$, whereas it is $-1$ when $|t|>|s|$. By the index theorem, Eq.\ \eqref{eqn:winding.index.signed}, $\hat{H}_L(A,V)$ and $\hat{H}_R(A,V)$ each has zero index ($|s|>|t|$ case) or $+1$ index ($|t|>|s|$ case). Clearly the $n=0$ degree of freedom is a $\bullet$ zero mode. The total index (mod 2) in both cases is thus
\[
{\rm Ind}(H_{\rm int}(A,V,A,V;0,0,0))=1\mod 2.
\]
Now, as we turn on the interface terms $B_L,B_R$, the chiral symmetry is preserved. Since the index is stable under such finite-rank perturbations, we still have
\[
{\rm Ind}(H_{\rm SSH})\equiv{\rm Ind}(H_{\rm int}(A,V,A,V;B_L,B_R,0))=1\mod 2.
\]
We have proved:
\begin{prop}\label{prop:SSH}
For hopping amplitudes $s,t$ with $|s|\neq |t|$, and arbitrary interface hopping terms, the SSH interface Hamiltonian, Eq.\ \eqref{eqn:SSH.model.Hamiltonian}, has a bulk spectral gap with an odd number of zero modes. 
\end{prop}
The existence of such interface modes was discussed in the seminal paper \cite{ssh-79} with a passing mention of index theorems in a subsequent work \cite{js-81}. To our knowledge, our proof of Prop.\ \ref{prop:SSH} is the first direct index-theoretic one.

\subsection{Interface modes in dislocation model}
While SSH model Hamiltonians exhibit a very clean formulation of bulk-interface correspondence, the reliance on strict chiral symmetry is problematic when we wish to model realistic continuum systems such as the photonic systems of Sec \ref{sec:inv_sym}, which only have $\mathsf{P}$ and $\mathsf{T}$ symmetry but no analogue of chiral symmetry.

In general, having $(A_L,V_L)=(A_R,V_R)=(A,V)$ in the interface model means that we put the ``same'' system on the left and right sides, but separate them by an extra half unit cell defect. Thus, we call $H_{\rm int}(A,V,A,V;B_L,B_R,W)$ a \textbf{dislocation model Hamiltonian}. \emph{With respect to a common origin}, the Zak phases for the right and left systems will therefore differ by $\pi$, due to Lemma \ref{lem:phase.shift}. Informally, we say that a ``trivial'' system has been placed next to a ``topological'' system.  

As in Theorem \ref{thm:PT.BEC}, let us extract the strictly nearest-neighbour part $H_{\rm int}^{\rm SNN}(A,V)$. It has bulk part, $H^{\rm SNN}(A,V)$, having the form of Eq.\ \eqref{eqn:SSN.Hamiltonian}, and the spectral gap is easily seen to be $(v_0-\Delta,v_0+\Delta)$, with
\[
\Delta=\big||s|-|t|\big|.
\]
Assuming that $H^{\rm far}(A,V)=H(A,V)-H^{\rm SNN}(A,V)$ has norm smaller than $\frac{\Delta}{2}$, its restoration does not close the gap, and the Zak phases of $H(A,V)$ and $H^{\rm SNN}(A,V)$ will be the same. We are interested in the relation between this Zak phase and the in-gap eigenvalues of $H_{\rm int}$.

\subsubsection*{Case where Zak$(H(A,V))=0$.} This occurs when $|s|>|t|$. Pictorially, 
\begin{alignat}{8}
\cdots& \;\;\;\;\circ &|&\;\;\;\; \bullet\,\circ \;\;\;\;&|&\;\;\;\; \bullet\,\circ \;\;\;\;&|& \;\bullet \;&|& \;\;\;\;\circ\,\bullet\;\;\;\; &|& \;\;\;\; \circ\,\bullet \;\;\;\;&|& \;\;\;\;\circ\,\bullet \;\;\;\;&|&\cdots\label{eqn:dislocated.lattice.1}\\
\cdots& -3 \;\;\;&|& \;\;-2 &|& \;\;-1 &|& \;0 &|& \;\;\;\;\;\;\;\; 1 &|& \;\;\;\;\;\;\;\; 2 &|& \;\;\;\;\;\;\;\;3 &|&\cdots\nonumber
\end{alignat}
with the two points in a unit cell being closer to each other, indicating that the intracell coupling is stronger than the intercell one. Ignoring interface terms, we have a direct sum
\[
H_{\rm int}(A,V,A,V;0,0,v_0)=\hat{H}_L(A,V)\oplus v_0\oplus \hat{H}_R(A,V).
\]
Note that the middle term is $v_0$, and we consider only the offset potential $W-v_0$ as part of the interface terms. 

\emph{Assume that $||H^{\rm far}(A,V)||_{\rm op}<\frac{\Delta}{2}$}. By Theorem \ref{thm:PT.BEC}, both $\hat{H}_L(A,V)$ and $\hat{H}_R(A,V)$ retain a spectral gap $(v_0-\frac{\Delta}{2},v_0+\frac{\Delta}{2})$ even after the $\hat{H}^{\rm far}_L(A,V), \hat{H}^{\rm far}_R(A,V)$ terms are restored. So $H_{\rm int}(A,V,A,V;0,0,v_0)$ has $v_0$ as an eigenvalue, spectrally isolated by a distance $\frac{\Delta}{2}$. This in-gap eigenvalue will survive the reintroduction of the interface terms, \emph{provided they are smaller than $\frac{\Delta}{4}$ in norm}. For example, one typically chooses
\begin{equation*}
B_R=\begin{pmatrix} t & 0 \end{pmatrix},\;\; B_L^*=\begin{pmatrix} 0 & t \end{pmatrix},\qquad \;\;W-v_0\ll \Delta.
\end{equation*}
Since $|t|<|s|$ and $\Delta=|s|-|t|$, it is possible to satisfy the small interface term condition.

\subsubsection*{Case where Zak$(H(A,V))=\pi$.} This occurs when $|t|>|s|$, i.e., the intercell hopping dominates the intracell one. Pictorially, instead of Eq.\ \eqref{eqn:dislocated.lattice.1}, we have
\begin{alignat}{7}
\cdots& \circ\, |\;\bullet\;\;\; &&\;\;\; \circ\,|\;\bullet\;\;\; && \;\;\;\circ\,|\,\;&\bullet&\;|\;\circ\;\;\; && \;\;\;\bullet\,|\;\circ\;\;\; && \;\;\;\bullet\,|\;\circ\;\;\; && \;\;\;\bullet\,|\;\;\;\cdots\nonumber\\
\cdots& -2\;\;\;\;\;\; &\;\;|&\;\;\;-1\;\;\;&\;\;|&\;\;\;\;\;\;\;\;&0&\;\;\;\;&\;\;|&\;\;\;\;\;\;\;\;1\;\;&\;\;|&\;\;\;\;\;\;\;\;2\;\;&\;\;|&\;\;\;\;\;\;\;\;3\;\;\;\cdots\nonumber
\end{alignat}
In the second line above, we introduce an alternative partition where the $n\neq 0$ unit cells are each shifted by half a unit cell. With this convention, the Zak phase of the left and right bulk systems becomes $0$, while the $n=0$ cell has \emph{three} degrees of freedom. 

First, we ignore the terms hopping in/out of $n=0$, so that the interface system can be separated into three independent parts. As before, $\hat{H}_L(A,V)$ and $\hat{H}_R(A,V)$ retain a spectral gap $(v_0-\frac{|t|-|s|}{2},v_0+\frac{|t|-|s|}{2})$, under the assumption that $H^{\rm far}(A,V)$ is smaller than $\frac{\Delta}{2}=\frac{|t|-|s|}{2}$. For the enlarged $n=0$ cell, the on-site term is
$
H_0=\begin{pmatrix} v_0 & t & 0 \\ t & v_0 & t \\ 0 & t & v_0 \end{pmatrix},
$
which has eigenvalues \mbox{$\{v_0-\sqrt{2}|t|,v_0,v_0+\sqrt{2}|t|\}$}. So \mbox{$\hat{H}_L(A,V)\oplus H_0\oplus \hat{H}_R(A,V)$} has $v_0$ as an in-gap eigenvalue, still isolated by a distance of at least $\frac{\Delta}{2}$. 

Now restore the interface term, which involves $s, B_R, B_L, W-v_0$ and other longer-range terms. Assuming this interface term has norm smaller than $\frac{\Delta}{4}$, the in-gap eigenvalue will survive. Note that $|s|<|t|$ and $\Delta=|t|-|s|$, so it is possible to satisfy the small interface term condition.

The results of this Subsection are summarized as follows:
\begin{thm}\label{thm:dislocated.bulk.interface.correspondence}
Let $H_{\rm int}=H_{\rm int}(A,V,A,V; B_L, B_R,W)$ be a dislocation model Hamiltonian, with $\mathsf{P}$ and $\mathsf{T}$ symmetric bulk Hamiltonian $H(A,V)$. Assume that for some overall scalar term $v_0$ and some $\Delta>0$, the strictly nearest-neighbour term $H^{\rm SSN}(A,V)$ has a spectral gap $(v_0-\Delta, v_0+\Delta)$, the term $H^{\rm far}(A,V)=H(A,V)-H^{\rm SNN}(A,V)$ has norm smaller than $\frac{\Delta}{2}$, and the interface term in $H_{\rm int}$ has norm smaller than $\frac{\Delta}{4}$. Then $H_{\rm int}$ has one in-gap interface mode inside the interval $(v_0-\frac{\Delta}{4},v_0+\frac{\Delta}{4})$.
\end{thm}



\begin{rem}
Whether $|s|>|t|$ or $|s|<|t|$ indicates two distinct ways of opening a spectral gap, starting from the equidistant case, $|s|=|t|$. In \cite{ssh-79,js-81}, this was called ``symmetry breaking'' induced by a staggered displacement field .
\end{rem}

\begin{rem}
We do not actually need $(A_L,V_L)=(A_R,V_R)$ in Theorem \ref{thm:dislocated.bulk.interface.correspondence}. As in Theorem \ref{thm-existence_int_mode}, we can allow $H(A_L,V_L)$ to be any $\mathsf{P}$ and $\mathsf{T}$ symmetric Hamiltonian with the same Zak phase as $H(A_R,V_R)$, as long as they share a common bulk gap, and the conditions on $H^{\rm SSN}$, $H^{\rm far}$ and interface terms are correspondingly made more conservative. 
\end{rem}

\section{Comparison of continuum and discrete models}\label{sec:comparison}
Continuum systems are often successfully modelled by finite-range lattice models, even if a full first-principles justification is seldom available. Famous lattice models, e.g. SSH models, employ further assumptions such as strictly nearest-neighbour interactions, to greatly simplify the analysis while still exhibiting interesting features (e.g., interface modes). However, when we go further and make statements about entire \emph{classes} of (e.g.\ interface, dislocation) lattice models constrained only by symmetries such as $\mathsf{P}$, $\mathsf{T}$, the link to realistic continuum models (e.g., \cite{ammari-20-1,ammari-20-3}) may become weakened.

For example, a ``purely topological'' version of Theorem \ref{thm:dislocated.bulk.interface.correspondence} might read: ``the interface of a trivial and topological phase has an in-gap interface mode''. This is false: Consider Example \ref{ex:simplest.interface}, which has bulk gap $(-1,1)$ and a zero-mode supported at $n=0$. Increasing the interface potential term $W$ pushes the in-gap eigenvalue into the bulk spectrum. 

Comparison with Prop.\ \ref{prop:dislocation.mode} is instructive. In the continuum model, the lowest band is known to have minimal energy at $k=0$ and even parity there. So when the lowest two bands are isolated from the others, the parity at $k=\pi$ suffices to determine the Zak phase of the lowest band (thus the bulk index $\gamma_1$). Prop.\ \ref{prop:dislocation.mode} and Theorem \ref{thm:dislocated.bulk.interface.correspondence} become very similar, but they differ in range of validity. For the lattice model, it will be interesting to analyze the conditions under which the in-gap interface mode of Theorem \ref{thm:dislocated.bulk.interface.correspondence} is the only one inside the entire bulk spectral gap, as is the case for the continuum model.

\begin{acknowledgments}
G.C.T. thanks K.~Yamamoto for helpful discussions on the SSH model. H.Z. is partially supported by the Hong Kong RGC grant GRF 16304621. 
\end{acknowledgments}


\end{document}